\PassOptionsToPackage{prologue,dvipsnames,svgnames}{xcolor}

\documentclass[10pt,twocolumn,twoside]{IEEEtran}
\usepackage{cite}
\usepackage{amsmath,amssymb,amsfonts}
\usepackage{algorithm}
\usepackage{algpseudocode}
\usepackage{graphicx}
\usepackage{textcomp}
\usepackage{amsthm}
\usepackage{kotex}
\usepackage[dvipsnames,svgnames]{xcolor}
\usepackage{array}
\usepackage{comment}
\usepackage{url}
\usepackage{matlab-prettifier}
\usepackage{listings}
\usepackage{hyperref}
\usepackage{siunitx}

\newtheorem{remark}{Remark}

\newcommand{\bbZ}{\ensuremath{{\mathbb Z}}}
\newcommand{\bbN}{\ensuremath{{\mathbb N}}}
\newcommand{\bbR}{\ensuremath{{\mathbb R}}}

\newcommand{\calT}{\ensuremath{\mathcal{T}}}
\newcommand{\calO}{\mathcal{O}}
\newcommand{\calC}{{\mathcal C}}

\newcommand{\sk}{{\mathrm{sk}}}
\newcommand{\Dec}{\mathsf{Dec}}
\newcommand{\Enc}{{\mathsf{Enc}}}
\newcommand{\Mult}{\mathsf{Mult}}
\newcommand{\Pack}{\mathsf{Pack}}
\newcommand{\UnpackPt}{\mathsf{UnpackPt}}
\newcommand{\UnpackCt}{\mathsf{UnpackCt}}
\newcommand{\Unpack}{\mathsf{Unpack}}
\newcommand{\coeff}{\mathsf{coeff}}
\newcommand{\ntt}{\mathsf{ntt}}

\newcommand{\sfs}{\mathsf{s}}
\newcommand{\sfL}{\mathsf{L}}
\newcommand{\sfr}{\mathsf{r}}

\newcommand{\ini}{\mathsf{ini}}

\newcommand{\bfc}{\mathbf{c}}
\newcommand{\bfF}{\mathbf{F}}
\newcommand{\bfG}{\mathbf{G}}
\newcommand{\bfH}{\mathbf{H}}
\newcommand{\bfR}{\mathbf{R}}
\newcommand{\bfx}{\mathbf{x}}
\newcommand{\bfy}{\mathbf{y}}
\newcommand{\bfu}{\mathbf{u}}

\DeclareMathOperator{\vect}{vec}
\newcommand{\col}{\ensuremath{\text{col}}}

\definecolor{lstgreen}{rgb}{0,0.5,0}

\def\BibTeX{{\rm B\kern-.05em{\sc i\kern-.025em b}\kern-.08em
    T\kern-.1667em\lower.7ex\hbox{E}\kern-.125emX}}

\makeatletter
\def\endthebibliography{%
	\def\@noitemerr{\@latex@warning{Empty `thebibliography' environment}}%
	\endlist
}
\makeatother

\begin{document}
	
\bstctlcite{IEEEexample:BSTcontrol}

\title{\LARGE \bf Documentation on Encrypted Dynamic Control\\
	Simulation Code using Ring-LWE based Cryptosystems}
\author{Yeongjun Jang, Joowon Lee, and Junsoo Kim
\thanks{This work was supported by the National Research Foundation of Korea(NRF) grant funded by the Korea government(MSIT) (No. RS-2024-00353032).}
\thanks{Y.~Jang and J.~Lee are with ASRI, the Department of Electrical and Computer Engineering, Seoul National University, South Korea (e-mail: \{jangyj, jwlee\}@cdsl.kr).}
\thanks{J.~Kim is with the Department of Electrical and Information Engineering,
	Seoul National University of Science and Technology, South Korea (e-mail: junsookim@seoultech.ac.kr).}  
}

\maketitle

\begin{abstract}
Encrypted controllers offer secure computation by employing modern cryptosystems to execute control operations directly over encrypted data without decryption. 
However, incorporating cryptosystems into dynamic controllers significantly increases the computational load. 
This paper aims to provide an accessible guideline for running encrypted controllers using an open-source library Lattigo, which supports an efficient implementation of Ring-Learing With Errors (LWE) based encrypted controllers, and our explanations are assisted with example codes that are fully available at \url{https://github.com/CDSL-EncryptedControl/CDSL}. 
\end{abstract}


\section{Introduction}\label{sec:intro}

Networked control is to outsource computations of physical devices to cloud servers in order to achieve computational efficiency and higher performance.
However, it also introduces the risk of cyber-attacks, as data that are transmitted and processed via public network are vulnerable to eavesdropping and manipulation by unauthorized third parties.
Encrypted control addresses this issue by employing homomorphic cryptosystems to perform control computations over encrypted data without decryption \cite{kogiso,csm,arc}.
This ensures the confidentiality of all data throughout the transmission and computation stages.

While significant theoretical advancements have been made in encrypted control, applying it to real-time systems remains challenging, primarily due to the high computational complexity of cryptosystems.
To ensure that all computations are completed within the sampling period, results found in the literature are typically limited to simple systems or compromise the level of security as a trade-off for reducing the computational burden, as in \cite{KimTAC,ecc}.

This issue could be relieved by utilizing some library implementations, such as Microsoft SEAL \cite{seal}, HElib \cite{helib}, and Lattigo \cite{lattigo}; 
they support optimized implementations of Ring-Learning With Errors (LWE) based cryptosystems \cite{rlwe,bfv,bgv,ckks}, which are among the most efficient and advanced homomorphic cryptosystem available.
However, such libraries often lack instructions for those without a strong cryptographic background, and thus, have not been actively used in the field of encrypted control.

This paper aims to provide an accessible guideline for implementing Ring-LWE based encrypted controllers using the Lattigo library, which supports an efficient implementation of state-of-the-art homomorphic cryptosystems.
In particular, we focus on the implementations of the encrypted controllers presented in \cite{Jang} and \cite{Lee}.
We briefly review each result, and provide some example codes, fully available at \url{https://github.com/CDSL-EncryptedControl/CDSL}, to assist our explanations.

The remainder of this paper is organized as follows.
Section~\ref{sec:rlwe} briefly introduces Ring-LWE based cryptosystems.
Sections~\ref{sec:TCNS} and~\ref{sec:TSMC} provide code-based explanations on implementing the encrypted controllers presented in \cite{Jang} and \cite{Lee}, respectively. Section~\ref{sec:conclusion} concludes the paper.

\textit{Notation:}
The set of integers, nonnegative integers, positive integers, and real numbers are denoted by $\bbZ$, $\bbZ_{\ge 0}$, $\bbN$, and $\bbR$, respectively. The Kronecker product and the Hadamard product (element-wise multiplication of vectors) are written by $\otimes$ and $\circ$, respectively.
The Moore-Penrose inverse and the vectorization of a matrix $A$ are denoted by $A^\dagger$ and $\vect(A)$, respectively.
For scalars $a_1,\,\ldots,\,a_n$, we define $\col\lbrace a_i\rbrace_{i=1}^n:=[a_1,\,\ldots,\,a_n]^\top$.
The vector $\mathbf{1}_n$ is defined as $[1,\,\ldots,\,1]^\top\in\bbR^n$.

\section{Ring-LWE based Cryptosystem}\label{sec:rlwe}

This section provides a brief overview of Ring-LWE based cryptosystems \cite{rlwe,bgv,bfv,ckks,rgsw}.
Ring-LWE based cryptosystems are based on the polynomial ring $R_q := \bbZ_q[X]/\langle X^N+1\rangle$, which is defined by parameters $q\in\bbN$ and $N\in\bbN$, where $N$ is a power of two.
The elements of $R_q$ can be represented by polynomials with indeterminate $X$, coefficients in $\bbZ_q:=\bbZ \cap [-q/2,q/2)$, and degree less than $N$.
The ring $R_q$ is closed under modular addition and multiplication, which regards $X^N\equiv -1$ and map each coefficient to the set $\bbZ_q$ via the modulo operation defined by $a\!\!\mod q:= a- \lfloor (a+q/2)/q\rfloor q$ for all $a\in\bbZ$.
The $l_\infty$ norm of a polynomial $m=\sum_{i=0}^{N-1}m_iX^i\in R_q$ is defined as $\lVert m \rVert:=\max_{0\leq i<N}\lvert m_i\rvert$.

In what follows, the setup, encryption and decryption algorithms, and homomorphic addition are described.
\begin{itemize}
	\item \textit{Parameters} $(N,q,\sigma)$:
	The degree $N\in\bbN$ is a power of two and
	the modulus $q\in\bbN$ is a prime satisfying $1=q\!\!\mod 2N$.
	The parameter $\sigma$ gives a bound on the error distribution.
	\item \textit{Key generation}: Generate the secret key $\sk\in R_q$.
	\item \textit{Encryption}: $\Enc:R_q\to R_q^2$
	\item \textit{Decryption}: $\Dec:R_q^2\to R_q$
	\item \textit{Homomorphic addition}: $\oplus:R_q^2\times R_q^2\to R_q^2$
\end{itemize}

The encryption and decryption algorithms satisfy the \textit{correctness} property, i.e., for any $m\in R_q$,
\begin{equation}\label{eq:correct}
	\Dec(\Enc(m))=m+e \!\!\! \mod q,
\end{equation}
where $e\in R_q$ is the error polynomial injected during the encryption of the message $m$ bounded by $\| e \|\le \sigma$.
The scheme also satisfies the \textit{additively homomorphic} property, written as
\begin{equation*}
	\Dec(\bfc_1 \oplus \bfc_2) = \Dec(\bfc_1) + \Dec(\bfc_2) \!\! \mod q,
\end{equation*}
for any $\bfc_1\in R_q^2$ and $\bfc_2\in R_q^2$.
The \textit{multiplicatively homomorphic} property will be described in Section~\ref{sec:TSMC}.

In \cite{Jang}, a method called ``external product'' \cite{rgsw} has been used to perform multiplications over encrypted data.
The external product is a product between a \textit{ciphertext} (encrypted data) in $R_q^2$ and another type of ciphertext generated by the Ring-GSW (Gentry-Sahai-Waters) encryption algorithm \cite{rgsw},
which is described below.
\begin{itemize}
	\item \textit{Parameters} $(d,\nu,P)$: Choose $d\in\bbN$ and a power of two $\nu\in\bbN$ such that $\nu^{d-1}<q\le \nu^d$, and the \textit{special modulus} $P\in\bbN$ as a prime.
	\item \textit{Ring-GSW encryption}: $\Enc^\prime:R_q \to R_{qP}^{2\times 2d}$
	\item \textit{External product}: $\boxdot$: $R_{qP}^{2\times 2d} \times R_q^2 \to R_q^2$
\end{itemize}

For any $m\in R_q$ and $\bfc \in R_q^2$, it holds that
\begin{equation}\label{eq:external}
	\Dec(\Enc^\prime(m)\boxdot \bfc) = m\cdot \Dec(\bfc) + e \!\!\! \mod q,
\end{equation}
where $e\in R_q$ is a ``newborn'' error under the external product bounded by $\|e\|\le \sigma_\Mult:=P^{-1}dN\sigma \nu+(N+1)/2$.
The property \eqref{eq:external} enables the recursive multiplications on the encrypted state of the controller proposed in \cite{Jang}.
The special modulus $P$ is used to temporarily ``lift'' the modulus from $q$ to $qP$ during the execution of the external product, thus reducing the bound $\sigma_\Mult$.

For simplicity, we abuse notation and define $\Enc(\cdot)$, $\Dec(\cdot)$, and $\|\cdot\|$ component-wisely for vectors throughout the paper.

\section{Recursive Encrypted Controller}\label{sec:TCNS}

In this section, we illustrate the implementation of the encrypted controller proposed in \cite{Jang}, focusing on its application to a four-tank system \cite{fourtank}, using Lattigo.
The encrypted controller enables an unlimited number of recursive homomorphic multiplications (by an encrypted integer matrix) without bootstrapping, which typically incurs high computational cost.
The computation speed can be further accelerated by applying a ``coefficient-packing'' technique that encodes a vector into a polynomial.
Example codes and instructions are included to assist readers in applying Lattigo for different systems or approaches as they choose,
which are fully available at:
\url{https://github.com/CDSL-EncryptedControl/CDSL/tree/main/ctrRGSW}

The plant is a four-tank system \cite{fourtank} discretized using the sampling time $100$ ms, written as
\begin{equation}\label{eq:plant}
	\begin{split}
		x_p(t+1) &= Ax_p(t) + Bu(t), \quad x_p(0) = x_p^\ini, \\ 
		y(t) &= Cx_p(t), 
	\end{split}
\end{equation}
where
\begin{equation*}
	\begin{split}
		A\!&=\!\!{\footnotesize\begin{bmatrix}
				0.9984 &\! 0 &\! 0.0042 &\! 0   \\
				0 &\! 0.9989 &\! 0 &\! -0.0033  \\
				0 &\! 0 &\! 0.9958 &\! 0 \\
				0 &\! 0  &\! 0  &\! 0.9967
		\end{bmatrix}}\!, \ 
		B \!=\!\! {\footnotesize\begin{bmatrix}
				0.0083 &\! 0 \\
				0 &\! 0.0063 \\
				0 &\! 0.0048 \\
				0.0031 &\! 0
		\end{bmatrix}}\!, \\ 
		C \!&=\!\! {\footnotesize\begin{bmatrix}
				0.5 & 0 & 0  & 0      \\
				0 & 0.5 & 0  & 0
		\end{bmatrix}}.
	\end{split}
\end{equation*}
With $n=4$, $m=2$, and $l=2$, $x_p(t)\in\bbR^{n}$ is the state with the initial value $x_p^{\ini}\in\bbR^{n}$, $u(t)\in\bbR^m$ is the input, and $y(t)\in\bbR^l$ is the output.

An observer-based controller that stabilizes \eqref{eq:plant} is designed as
\begin{equation}\label{eq:controller}
	\begin{split}
		x(t+1) &= Fx(t) + Gy(t), \quad x(0) = x^\ini, \\ 
		u(t) &= Hx(t), 
	\end{split}
\end{equation}
where $x(t)\in\bbR^n$ is the state with the initial value $x^\ini\in\bbR^n$, and
the matrices are given by $F=A+BK-LC$, $G=L$, and $H=K$ with
\begin{equation*}
	\begin{split} 
		K& = {\footnotesize\begin{bmatrix}
				-0.7905 & 0.1579 & -0.2745 & -0.2686 \\
				-0.1552 & -0.7874 & -0.3427 & 0.3137
		\end{bmatrix}},  \\
		L & = {\footnotesize\begin{bmatrix}
				0.7815 & 0 & 0.3190 & 0 \\
				0 & 0.7816 & 0 & -0.3199
		\end{bmatrix}}^\top. 
	\end{split}
\end{equation*}
It can be easily verified that \eqref{eq:controller} is controllable and observable.

The objective is to design an encrypted controller that performs the operations of \eqref{eq:controller} over encrypted data exploiting the homomorphic properties described in Section~\ref{sec:rlwe}.
The design should guarantee that the performance of the encrypted controller can be made arbitrarily close to that of the unencrypted controller \eqref{eq:controller} with an appropriate choice of parameters.

\subsection{Encryption setting}

In Lattigo, one way to setup the parameters of Ring-LWE based cryptosystems is as follows:
\begin{lstlisting}[style=Matlab-editor,
	basicstyle=\ttfamily\footnotesize,
	commentstyle=\color{lstgreen}\ttfamily,
	mathescape=true] 
params, _ := rlwe.NewParametersFromLiteral(
rlwe.ParametersLiteral{
	LogN: 13,          % polynomial degree
	LogQ: []int{56},   % ciphertext modulus $\color{lstgreen}q$
	LogP: []int{51}})  % special modulus $\color{lstgreen}P$
\end{lstlisting}
The above code sets $N=2^{13}$, and automatically selects primes $q\approx2^{56}$ and $P\approx 2^{51}$ meeting the requirement $1=q\!\!\mod 2N$, which are $q=72057594037616641$ and $P=2251799813554177$ in this example.
The chosen parameters ensure the 128-bit security according to \cite{HEstandard}.
By default, the error distribution is defined as a discrete zero-mean Gaussian distribution with standard deviation $3.2$ and bound $\sigma = 19.2$, and the parameters for the Ring-GSW encryption are set as $\nu=q$ and $d=1$.

Given the parameters, we generate the secret key using
\begin{lstlisting}[style=Matlab-editor,
	basicstyle=\ttfamily\footnotesize,
	commentstyle=\color{lstgreen}\ttfamily,
	mathescape=true] 
kgen := rlwe.NewKeyGenerator(params)
sk := kgen.GenSecretKeyNew() % secret key $\color{lstgreen}\sk\in R_q$
\end{lstlisting}
where each coefficient of $\sk$ is sampled uniformly at random from the set $\{-1,0,1\}$ by default.

Next, we briefly describe the packing technique proposed in \cite{Jang}.
Let us define $\tau\in\bbN$ as the smallest power of two satisfying $\tau\ge \max\{n,m,l\}$.
For any $k\in\bbN$ such that $k\le \tau$, the packing algorithm $\Pack^k_\coeff:\bbZ_q^k \to R_q$ encodes each component of a vector as coefficients of the terms $X^0,X^{N/\tau},\ldots,X^{(k-1)N/\tau}$ in sequential order, and the corresponding unpacking algorithm $\UnpackPt^k_\coeff:R_q \to \bbZ_q^k$ reconstructs the original vector, i.e., 
\begin{equation*}
	\UnpackPt^k_\coeff(\Pack^k_\coeff(a)) = a, \quad \forall a\in\bbZ_q^k.
\end{equation*}

In fact, an algorithm $\UnpackCt^k_\coeff:R_q^2 \to R_q^{2k}$ that homomorphically evaluates $\UnpackPt^k_\coeff$ is proposed in \cite{Jang},
which satisfies
\begin{equation*}
	\UnpackPt^k_\coeff(\Dec(\bfc)) = \Dec(\UnpackCt^k_\coeff(\bfc)) + e
\end{equation*}
for any $\bfc\in R_q^2$,
with some $e\in R_q^k$ bounded by $\| e\|\le \sigma_\Mult\log_2\tau$.
Notably, this allows efficient matrix-vector multiplications over encrypted data; see \cite[Section~IV]{Jang} for more details.

To make use of the unpacking algorithm $\UnpackCt^k_\coeff$, we need to define some additional \textit{evaluation keys} as follows:
\begin{lstlisting}[style=Matlab-editor,
	basicstyle=\ttfamily\footnotesize,
	mathescape=true] 
rlk := kgen.GenRelinearizationKeyNew(sk)
evkRGSW := rlwe.NewMemEvaluationKeySet(rlk)
evkRLWE := rlwe.NewMemEvaluationKeySet(rlk,
	kgen.GenGaloisKeysNew(galEls, sk)$\ldots$)
\end{lstlisting}

\subsection{Encrypted controller design \cite{Jang}}

It is known that the state matrix of the controller needs to consist of integers for a dynamic system to operate on encrypted data \cite{integer}.
The code \verb|ctrRGSW/conversion.m| converts the state matrix $F$ of \eqref{eq:controller} into integers based on the approach of \cite{KimTAC}, which ``re-encrypts'' the output:
we reformulate the state dynamics of \eqref{eq:controller}, as
\begin{equation*}
	x(t+1) = (F-RH)x(t) + Gy(t) + Ru(t),
\end{equation*}
where $R\in \bbR^{n\times m}$ is designed as
\begin{equation*}
	R = {\footnotesize  \begin{bmatrix}
			-1.6879 & 0.4148 & 0.2880 & -0.7385 \\
			-0.6892 & -2.1054 & 4.1931 & -2.0807
		\end{bmatrix}^\top}, 
\end{equation*}
so that $\bar{F}:=(F-RH)\in \bbZ^{n\times n}$.
We now consider $\bar{F}$ as the state matrix of \eqref{eq:controller} regarding $u(t)$ as a fed-back input.

In \verb|ctrRGSW/packing/main.go|,
each column of the control parameters $\{F,G,H,R\}$ is quantized using a scale factor $1/\sfs\in\bbN$, packed, and then encrypted, as
\begin{align*}
	\bfF_i &:= \Enc'(\Pack^n_\coeff(F_i\!\!\!\mod q)),& i&=0,\ldots,n-1, \\
	\bfG_i &:= \Enc'(\Pack^n_\coeff(\lceil G_i/\sfs \rfloor \!\!\!\mod q)),& i&=0,\ldots,l-1,\\
	\bfH_i &:= \Enc'(\Pack^m_\coeff(\lceil H_i/\sfs \rfloor \!\!\!\mod q)),& i&=0,\ldots,n-1, \\
	\bfR_i &:= \Enc'(\Pack^n_\coeff(\lceil R_i/\sfs \rfloor \!\!\!\mod q)),& i&=0,\ldots,m-1,
\end{align*}
where $\{F_i,G_i,H_i,R_i\}$ denote the $(i+1)$-th column of $\{F,G,H,R\}$.
This can be realized in the case of $G$ as
\begin{lstlisting}[style=Matlab-editor,
	basicstyle=\ttfamily\footnotesize,
	commentstyle=\color{lstgreen}\ttfamily,
	mathescape=true] 
GBar := utils.ScalMatMult(1/s, G) % scale up
ctG := RGSW.EncPack(Gbar, tau, encryptorRGSW,   
  levelQ, levelP, ringQ, params) % pack and encrypt
\end{lstlisting}
Here, the variable \verb|ctG| represents a $l$-dimensional vector of Ring-GSW type ciphertexts $\bfG_i$'s.

Similarly, the initial state $x^\ini$ is quantized, packed, and encrypted as 
\begin{equation*}
	\bfx^\ini := \Enc(\Pack^n_\coeff(\lceil x^\ini/(\sfr\sfs)\rfloor/\sfL  \!\!\! \mod q)), 
\end{equation*}
which is realized via
\begin{lstlisting}[style=Matlab-editor,
	basicstyle=\ttfamily\footnotesize,
	commentstyle=\color{lstgreen}\ttfamily,
	mathescape=true] 
xBar := utils.ScalVecMult(1/(r*s),x_ini)
xCtPack := RLWE.EncPack(xScale, tau, 1/L, 
	*encryptorRLWE, ringQ, params)
\end{lstlisting}
The parameter $\sfr>0$ represents the quantization step size of the sensor, and the additional scale factor $1/\sfL\in\bbN$ is introduced to reduce the effect of errors that can be found in \eqref{eq:correct} and \eqref{eq:external}, for example.
The parameters $\lbrace \sfs,\sfL, \sfr\rbrace$ are set as
\begin{lstlisting}[style=Matlab-editor,
	basicstyle=\ttfamily\footnotesize,
	mathescape=true] 
s := 1/10000.0    L := 1/10000.0    r := 1/10000.0
\end{lstlisting}

At each time step $t\in\bbZ_{\ge 0}$ of the online control procedure, the plant output $y(t)$ is encrypted at the sensor, as
\begin{equation*}
	\bfy(t) := \Enc(\Pack^l_\coeff(\lceil y(t)/\sfr \rfloor / \sfL \!\! \mod q)).
\end{equation*}
Then, it is sent to the encrypted controller, written by
\begin{align}\label{eq:encController}
	\bfx(t+1) &= \left(\sum_{i=0}^{n-1} \bfF_i\boxdot\bfx_i(t) \right) \oplus \left( \sum_{i=0}^{l-1}\bfG_i\boxdot\bfy_i(t) \right) \nonumber\\
	&\quad \oplus \left( \sum_{i=0}^{m-1}\bfR_i\boxdot\bfu_i(t) \right),  \\
	\bfu(t) &= \sum_{i=0}^{n-1} \bfH_i\boxdot\bfx_i(t),\qquad
	\bfx(0) = \bfx^\ini, \nonumber
\end{align}
where we let the output $\bfu(t)$ be decrypted at the actuator to obtain the plant input $u(t)$ and the re-encrypted fed-back inputs $\bfu_i(t)$'s, as 
\begin{equation*}
	\begin{split}
		&u(t) = \sfr\sfs^2\sfL\cdot \UnpackPt^m_\coeff(\Dec(\bfu(t))), \\
		&[\bfu_0(t);\cdots;\bfu_{m-1}(t)]:=\Enc(\lceil u(t)/\sfr \rfloor/\sfL \!\!\mod q),
	\end{split}
\end{equation*}
and $\bfx_i(t)$'s and $\bfy_i(t)$'s are obtained through unpacking:
\begin{equation*}
	\begin{split}
		[\bfx_0(t);\cdots;\bfx_{n-1}(t)]&:=\UnpackCt^{n}_\coeff(\bfx(t)), \\
		[\bfy_0(t);\cdots;\bfy_{l-1}(t)]&:=\UnpackCt^{l}_\coeff(\bfy(t)).
	\end{split}
\end{equation*}

The overall procedure at each time step can be simulated using the following code:
\begin{lstlisting}[style=Matlab-editor,
	basicstyle=\ttfamily\footnotesize,
	commentstyle=\color{lstgreen}\ttfamily,
	mathescape=true] 
y := utils.MatVecMult(C, xp)    % @ plant
yBar := utils.RoundVec(utils.ScalVecMult(1/r,y))			    % @ sensor
yCtPack := RLWE.EncPack(yBar, tau, 1/L, 
 *encryptorRLWE, ringQ, params) % @ sensor
xCt := RLWE.UnpackCt(xCtPack, n, tau, evaluatorRLWE, 
  ringQ, monomials, params) 	% @ controller
yCt := RLWE.UnpackCt(yCtPack, p, tau, evaluatorRLWE, 
  ringQ, monomials, params) 	% @ controller
uCtPack := RGSW.MultPack(xCt, ctH, evaluatorRGSW, 
	ringQ, params) 		% @ controller 
u := RLWE.DecUnpack(uCtPack, m, tau, *decryptorRLWE, 
	r*s*s*L, ringQ, params) % @ actuator  
uBar := utils.RoundVec(utils.ScalVecMult(1/r, u)) 
				% @ actuator  
uReEnc := RLWE.Enc(uBar, 1/L, *encryptorRLWE, 
	ringQ, params) 		% @ actuator  
FxCt := RGSW.MultPack(xCt, ctF, evaluatorRGSW, 
	ringQ, params) 		% @ controller
GyCt := RGSW.MultPack(yCt, ctG, evaluatorRGSW, 
	ringQ, params) 		% @ controller
RuCt := RGSW.MultPack(uReEnc, ctR, evaluatorRGSW, 
	ringQ, params) 		% @ controller
xCtPack = RLWE.Add(FxCt, GyCt, RuCt, params)
				% @ controller
xp = utils.VecAdd(utils.MatVecMult(A, xp), 
	utils.MatVecMult(B, u)) % @ plant
\end{lstlisting}

We define the performance error as the difference between the plant input generated by the unencrypted controller \eqref{eq:controller} and the encrypted controller \eqref{eq:encController}.
In this example, the maximum and average of the performance error for $1000$ iterations are obtained as $0.0089$ and $0.0028$, respectively.
Also, it took $\SI{16.6}{\milli\s}$ on average to compute the control input at each time step, which falls within the sampling time.
The encrypted controller implemented without packing can be found in the code \verb|ctrRGSW/noPacking/main.go|, which exhibits a slower operation.

\begin{remark}\upshape
	To decrease the performance error, the parameters $\lbrace \sfs,\sfL, \sfr \rbrace$ can be decreased.
	However, this may lead to overflow, where the encrypted messages grow out of $\bbZ_q$, resulting in controller malfunctions.
    The modulus $q$ can be increased to avoid such overflow, but greater $q$ often leads to greater $N$ under a fixed level of security \cite{HEstandard}, which significantly affects the computation time.
    Another way to decrease the performance error is to increase $P$, but again, this may require larger $N$.
\end{remark}

\section{Non-recursive Encrypted Controller}\label{sec:TSMC}

This section presents an alternative method \cite{Lee} to encrypt a dynamic controller using Ring-LWE based cryptosystems,
provided with a code implementation based on Lattigo.
This method
offers a faster computation and
can be applied to various Ring-LWE based schemes such as BGV \cite{bgv}, BFV \cite{bfv}, and CKKS \cite{ckks}.
However, unlike the method of Section~\ref{sec:TCNS}, it avoids recursive homomorphic multiplications,
and therefore inherently requires re-encryption of the controller output.

The implementation is again based on the four-tank system, where the plant and the controller are the same as \eqref{eq:plant} and \eqref{eq:controller}, respectively.
In this example, the BGV scheme \cite{bgv} is utilized, but the code can be easily modified for the use of other Ring-LWE based schemes.
On average, it takes less than $12$ ms for the resulting encrypted controller to compute the control input at each time step.
Readers can refer to the entire set of codes at: \url{https://github.com/CDSL-EncryptedControl/CDSL/tree/main/ctrRLWE}

\subsection{Encryption setting}

To begin with, we introduce another parameter for encryption, the \textit{plaintext} (message to be encrypted) modulus $p\in\bbN$, which is a prime satisfying $1=p\!\!\mod 2N$.
In the BGV scheme, the plaintext space is $R_p$, and the ciphertext space is $R_q^2$, where $q$ is referred as the ciphertext modulus.
Therefore, the encryption and decryption algorithms are defined as $\Enc:R_p\to R_q^2$ and $\Dec:R_q^2\to R_p$, respectively, throughout this section.
Under the assumption that $p\ll q$, the correctness property is rewritten as
\begin{equation*}
    \Dec(\Enc(m))=m\!\!\! \mod p,\quad \forall m\in\bbZ_p.
\end{equation*}

The method of \cite{Lee} utilizes homomorphic multiplication $\odot:R_q^2\times R_q^2\to R_q^3$,
which increases the dimension of ciphertexts by one.
Accordingly, the homomorphic addition $\oplus$ and decryption $\Dec(\cdot)$ are also defined over $R_q^3$.
The multiplicatively homomorphic property holds, i.e.,
\begin{equation*}
    \Dec(\Enc(m_1)\odot\Enc(m_2))=m_1\cdot m_2\!\!\!\mod p,
\end{equation*}
for any $m_1\in R_p$ and $m_2\in R_p$.

In this section, we exploit the packing method based on the number-theoretic transform (NTT) \cite{ntt}, which transforms a vector in $\bbZ_p^N$ into a polynomial in $R_p$ and vice versa.
Let the NTT-style packing and unpacking functions be denoted by $\Pack_{\ntt}:\bbZ_p^N\to R_p$ and $\Unpack_{\ntt}:R_p\to\bbZ_p^N$, respectively.
Then, the followings hold for any $a\in\bbZ^N_p$ and $b\in\bbZ^N_p$:
\begin{equation}\label{eq:ntt}
\begin{aligned}
    \Unpack_{\ntt}(\Pack_{\ntt}(a))&=a\!\! \mod p,\\
    \Unpack_{\ntt}(\Pack_{\ntt}(a)+\Pack_{\ntt}(b))&=a+b\!\!\mod p,\\
    \Unpack_{\ntt}(\Pack_{\ntt}(a)\cdot\Pack_{\ntt}(b))&=a\circ b\!\!\mod p.
\end{aligned}
\end{equation}
Thus, the element-wise additions and multiplications among vectors can be operated at once over encrypted data.
We abuse notation and apply $\Pack_{\ntt}$ to vectors with length shorter than $N$, regarding that zeros are ``padded'' at the end to fit the length.

In \verb|ctrRLWE/main.go|, the following code assigns the encryption parameters to be $N=2^{12}$, $p\approx 2^{26}$, and $q\approx 2^{74}$:
\begin{lstlisting}[style=Matlab-editor,
        basicstyle=\ttfamily\footnotesize,
        commentstyle=\color{lstgreen}\ttfamily,
        mathescape=true] 
logN:=12      ptSize:=uint64(28)    ctSize:=int(74)
\end{lstlisting}
The above parameters are chosen to meet the $128$-bit security, based on \cite{HEstandard}.
The code automatically finds a suitable prime $p$ that satisfies $1=p\!\!\mod 2N$,
saving the exact value of $p$ as \verb|ptModulus|, which is $p=268460033$ in this case.

To represent large integers in $\bbZ_q$, where $q\approx 2^{74}$, we create a chain of ciphertext moduli $\bbZ_{q^\prime}\times \bbZ_{q^\prime}$, where $q^\prime\approx 2^{37}$.
Thus, the code assigns $[37,\,37]$ to \verb|logQ|, from which Lattigo finds a proper prime $q^\prime$ by running the following code:
\begin{lstlisting}[style=Matlab-editor,
        basicstyle=\ttfamily\footnotesize,
        commentstyle=\color{lstgreen}\ttfamily,
        mathescape=true] 
params, _ := bgv.NewParametersFromLiteral(bgv.ParametersLiteral{
	LogN:             logN,
	LogQ:             logQ,
	PlaintextModulus: ptModulus,})
\end{lstlisting}

The NTT-style packing operation is enabled in Lattigo by generating a new ``encoder'' as follows:
\begin{lstlisting}[style=Matlab-editor,
        basicstyle=\ttfamily\footnotesize,
        commentstyle=\color{lstgreen}\ttfamily,
        mathescape=true] 
encoder := bgv.NewEncoder(params)
\end{lstlisting}
Meanwhile, the method of \cite{Lee}
only utilizes basic homomorphic addition and multiplication.
Therefore, the homomorphic evaluator is constructed simply as follows:
\begin{lstlisting}[style=Matlab-editor,
        basicstyle=\ttfamily\footnotesize,
        commentstyle=\color{lstgreen}\ttfamily,
        mathescape=true] 
eval := bgv.NewEvaluator(params, nil)
\end{lstlisting}

\subsection{Encrypted controller design \cite{Lee}}

The method of \cite{Lee} first converts the given controller \eqref{eq:controller} into an equivalent form;
\begin{equation}\label{eq:io form}
    u(t)=\sum_{i=1}^n H_{u,i}u(t-i)+H_{y,i}y(t-i),
\end{equation}
where $H_{u,i}\in\bbR^{m\times m}$ and $H_{y,i}\in\bbR^{m\times l}$ for $i=1,\,\ldots,\,n$.
The matrices in \eqref{eq:io form} are obtained as
\begin{align*}
    \left[H_{u,n},\,\ldots,\,H_{u,1}\right]&=HF^n\calO^\dagger,\\
    \left[H_{y,n},\,\ldots,\,H_{y,1}\right]&=H\left(\calC-F^n\calO^\dagger\calT\right),
\end{align*}
where
\begin{align*}
    \calO&={\small\begin{bmatrix}
        H\\ HF\\ \vdots \\ HF^{n-1}
    \end{bmatrix}},\,
    \calT = {\small\begin{bmatrix}
        0 & 0 & \cdots & 0\\
        HG & 0 & \cdots & 0\\
        \vdots & \ddots & \ddots & \vdots\\
        HF^{n-2}G & \cdots & HG & 0
    \end{bmatrix}},\\
    \calC &= {\small\begin{bmatrix}
        F^{n-1}G & \cdots & FG & G
    \end{bmatrix}},
\end{align*}
with $0$ denoting zero matrices of appropriate dimensions.
Furthermore,
to compute $u(t)$ from \eqref{eq:io form} for $t=0,\,\ldots,\,n-1$, we obtain the following thanks to \cite[Lemma 1]{Lee}:
\begin{equation}\label{eq:init yu}
\begin{aligned}
    \left[y(-n)^\top,\,\ldots,\,y(-1)^\top\right]^\top &=\calC^\dagger x^{\ini},\\
    \left[u(-n)^\top,\,\ldots,\,u(-1)^\top\right]^\top&=\calT\calC^\dagger x^{\ini}.
\end{aligned}
\end{equation}
This is implemented in \verb|ctrRLWE/conversion.m|.

To make use of the properties \eqref{eq:ntt}, the matrix-vector multiplications in \eqref{eq:io form} are reformulated, using the fact that
\begin{equation*}
    Ab = \col\left\lbrace \left\langle \vect(A^\top)\circ \left(\mathbf{1}_m\otimes b\right),e_i\otimes \mathbf{1}_h\right\rangle\right\rbrace_{i=1}^m
\end{equation*}
for any $A\in\bbR^{m\times h}$ and $b\in\bbR^h$,
where $e_i\in \bbR^m$ is the unit vector whose only nonzero element is its $i$-th element.

The matrices in \eqref{eq:io form} are vectorized, quantized, packed, and then encrypted as
\begin{align*}
    \bfH_{u,i}&:=\Enc(\Pack_{\ntt}(\left\lceil \vect(H_{u,i}^\top)/\sfs\right\rfloor\!\!\!\mod p)),\\
    \bfH_{y,i}&:=\Enc(\Pack_{\ntt}(\left\lceil \vect(H_{y,i}^\top)/\sfs\right\rfloor\!\!\!\mod p)),
\end{align*}
for $i=1,\,\ldots,\,n$.
In \verb|ctrRLWE/main.go|, $\bfH_{u,n-j}$ and $\bfH_{y,n-j}$ correspond to \verb|ctHu[j]| and \verb|ctHy[j]|, respectively, for $j=0,\,\ldots,\,n-1$.

The encrypted controller is designed as
\begin{equation}\label{eq:enc ctr}
    \bar{\bfu}(t)=\sum_{i=1}^n \bfH_{u,i}\odot\bfu(t-i)\oplus\bfH_{y,i}\odot\bfy(t-i),
\end{equation}
where $\bfy(t)$ and $\bfu(t)$ are inputs from the sensor and the actuator, respectively, defined as
\begin{equation}\label{eq:enc yu}
\begin{aligned}
    \bfy(t)&=\Enc(\Pack_{\ntt}(\left\lceil \mathbf{1}_m\otimes y(t)/\sfr\right\rfloor\!\!\!\mod p)),\\
    \bfu(t)&=\Enc(\Pack_{\ntt}(\left\lceil \mathbf{1}_m\otimes u(t)/\sfr\right\rfloor\!\!\!\mod p)).
\end{aligned}
\end{equation}
The initial input-output trajectories \eqref{eq:init yu} are encrypted as in \eqref{eq:enc yu} during offline.
At the actuator, the plant input is computed as
\begin{equation}\label{eq:ctr input}
    u(t)=\col\left\lbrace \left\langle \sfr \sfs\cdot\Unpack_{\ntt}(\Dec(\bar{\mathbf{u}}(t))),e_i\otimes \mathbf{1}_h\right\rangle\right\rbrace_{i=1}^m.
\end{equation}

In fact, this design requires that all of $\mathbf{1}_m\otimes y(t)$, $\mathbf{1}_m\otimes u(t)$, $\vect(H_{u,i}^\top)$'s, and $\vect(H_{y,i}^\top)$'s have the same length.
Thus, zeros are padded to ensure that they are all of length $mh$, where $h:=\max\lbrace m,\,l\rbrace$.
The function \verb|utils.VecDuplicate| is utilized to generate $\mathbf{1}_m\otimes y(t)$ and $\mathbf{1}_m\otimes u(t)$, with zero padding if necessary.
The vectorized control parameters with zero padding are precomputed in \verb|ctrRLWE/conversion.m|.

The Go code \verb|ctrRLWE/main.go| implements the encrypted controller \eqref{eq:enc ctr} as follows:
\begin{lstlisting}[style=Matlab-editor,
        basicstyle=\ttfamily\footnotesize,
        commentstyle=\color{lstgreen}\ttfamily,
        mathescape=true]
Uout, _ := eval.MulNew(ctHy[0], ctY[0])
eval.MulThenAdd(ctHu[0], ctU[0], Uout)
for j := 1; j < n; j++ {
    eval.MulThenAdd(ctHy[j], ctY[j], Uout)
    eval.MulThenAdd(ctHu[j], ctU[j], Uout)  }
\end{lstlisting}
Here, the function \verb|eval.MulThenAdd| homomorphically adds the first and the second arguments, assigning the outcome to the third argument.
The variables \verb|ctY[j]| and \verb|ctU[j]| correspond to $\bfy(t-n+j)$ and $\bfu(t-n+j)$, respectively, for $j=0,\,\ldots,\,n-1$.

The inputs \eqref{eq:enc yu} of the encrypted controller are generated as
\begin{lstlisting}[style=Matlab-editor,
        basicstyle=\ttfamily\footnotesize,
        commentstyle=\color{lstgreen}\ttfamily,
        mathescape=true] 
Ysens:=utils.ModVecFloat(utils.RoundVec(utils.ScalVecMult(1/r, utils.VecDuplicate(Y, m, h))), params.PlaintextModulus())
Ypacked:=bgv.NewPlaintext(params, params.MaxLevel())
encoder.Encode(Ysens, Ypacked)
Ycin, _:=encryptor.EncryptNew(Ypacked)
\end{lstlisting}
and
\begin{lstlisting}[style=Matlab-editor,
        basicstyle=\ttfamily\footnotesize,
        commentstyle=\color{lstgreen}\ttfamily,
        mathescape=true] 
Upacked:=bgv.NewPlaintext(params, params.MaxLevel())
encoder.Encode(utils.ModVecFloat(utils.RoundVec(utils.ScalVecMult(1/r, utils.VecDuplicate(U, m, h))), params.PlaintextModulus()), Upacked)
Ucin, _:=encryptor.EncryptNew(Upacked)
\end{lstlisting}
where \verb|Y| and \verb|U| refer to the plant output and input, respectively.

Finally, the operations at the actuator \eqref{eq:ctr input} are realized as follows:
\begin{lstlisting}[style=Matlab-editor,
        basicstyle=\ttfamily\footnotesize,
        commentstyle=\color{lstgreen}\ttfamily,
        mathescape=true]
Uact := make([]uint64, params.N())
Usum := make([]uint64, m)
encoder.Decode(decryptor.DecryptNew(Uout), Uact)
for k := 0; k < m; k++ {
    Usum[k] = utils.VecSumUint(Uact[k*h:(k+1)*h], params.PlaintextModulus(), bredparams)
    U[k] = float64(r * s * utils.SignFloat(float64(Usum[k]), params.PlaintextModulus()))    }
\end{lstlisting}
The variable \verb|bredparams| is for addition over $\bbZ_p$ in Lattigo.

In this example, the parameters $\sfr$ and $\sfs$ are set as $0.0002$ and $0.0001$, respectively.
The performance error is $0.00254$ on average, and $0.015788$ at maximum.
As mentioned in Remark~1, the parameters $\sfr$ and $\sfs$ can be decreased for better performance, however, one should be aware that the plaintext modulus $p$ needs to satisfy $\lVert u(t)\rVert_{\infty}/(\sfr\sfs)<p/2$, in order to avoid overflow.

\section{Conclusion}\label{sec:conclusion}

This paper provides example codes with instructions on implementing Ring-LWE based encrypted controllers \cite{Jang,Lee} for a four-tank system using Lattigo.
The codes can be easily customized by readers for application to different systems or approaches.
It is demonstrated that the methods of \cite{Jang} and \cite{Lee} enable fast implementations of encrypted controllers, taking less than $17$ ms on average at each time step for computing the control input of a fourth-order controller.




\bibliographystyle{IEEEtran}
\bibliography{sice}

%
%

\begin{IEEEbiography}[{\includegraphics[width=1in,height=1.25in,clip,keepaspectratio]{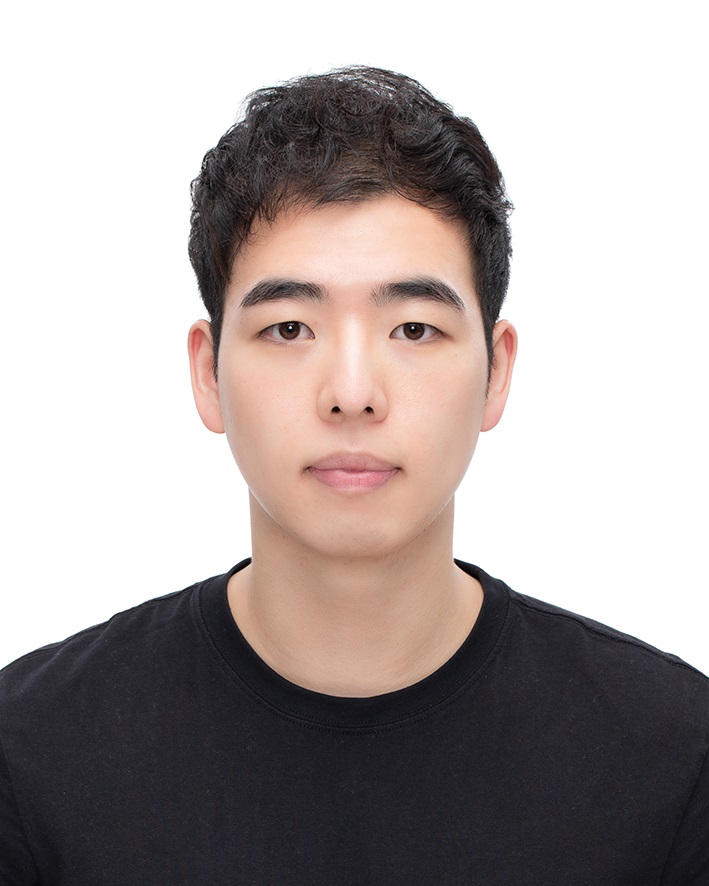}}]%
	{Yeongjun Jang}
	received the B.S. degree in electrical and computer engineering in 2022, from Seoul National University, South Korea.
	He is currently a combined M.S./Ph.D. student in electrical and computer engineering at Seoul National University, South Korea. 
	His research interests include data-driven control and encrypted control systems.
\end{IEEEbiography}

\begin{IEEEbiography}[{\includegraphics[width=1in,height=1.25in,clip,keepaspectratio]{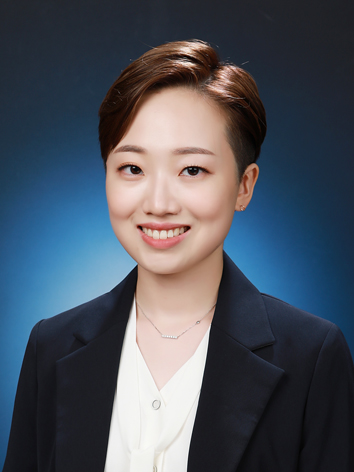}}]%
	{Joowon Lee}
	received the B.S. degree in electrical and computer engineering in 2019, from Seoul National University, South Korea.
	She is currently a combined M.S./Ph.D. student in electrical and computer engineering at Seoul National University, South Korea. 
	Her research interests include data-driven control, encrypted control systems, and cyber-physical systems.
\end{IEEEbiography}

\begin{IEEEbiography}[{\includegraphics[width=1in,height=1.25in,clip,keepaspectratio]{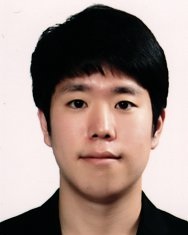}}]%
	{Junsoo Kim}
	received the B.S. degrees in electrical engineering and mathematical sciences in
	2014, and the M.S. and Ph.D. degrees in electrical engineering in 2020, from Seoul National University, South Korea, respectively. 
	He held the Postdoc position at KTH Royal Institute of Technology, Sweden, till 2022. 
	He is currently an
	Assistant Professor at the Department of Electrical and Information Engineering, Seoul National University of Science and Technology, South Korea. His research interests include security
	problems in networked control systems and encrypted control systems.
\end{IEEEbiography}

\end{document}